\documentclass[twoside]{article}
\usepackage{fleqn,espcrc2}

\usepackage{graphicx}
\usepackage[figuresright]{rotating}

\newcommand{\AmS}{{\protect\the\textfont2
  A\kern-.1667em\lower.5ex\hbox{M}\kern-.125emS}}

\title{Hidden Sector Supergravity Breakdown}

\author{Hans Peter Nilles
\address{Physikalisches Institut,  
        Universit\"at Bonn, 
        Nussallee 12, D-53115 Bonn, Germany}%
\thanks{Invited talk presented at the symposium ``Thirty Years
of Supersymmetry'', University of Minnesota, October 2000.
Work supported by grants of the European community
RTN programmes HPNR-CT-2000-00131, 00148 and 00152.}}
       
\begin{document}

\begin{abstract}

Gravity mediated supersymmetry breakdown originated from a study
of gaugino condensation in a hidden sector.
We review this mechanism of supergravity breakdown from 
the original formulation in the early eighties 
to its
natural implementation in superstring theory and M-theory. In the 
latter case hidden and observable sector become separated geometrically
and supersymmetry is broken dynamically on a hidden wall.

\vspace{1pc}
\end{abstract}

\maketitle

\section{Introduction}

In the early 1980's considerations of the hierarchy problem in the
standard model of strong and electroweak interactions (as well as grand
unified theories) strengthened the interest in applications of
supersymmetry to particle physics. I was at SLAC at that time and I
remember being intrigued by the idea that the scale of dynamically
broken supersymmetry \cite{Witten:1981nf}
would explain the weak mass scale of the standard
model and induce the breakdown of $SU(2)\times U(1)$. Still this goal was not
so easy to achieve. After the first excitement of model building in the
framework of global supersymmetry, problems started to appear. One of
them was the problem of the cosmological constant that was strictly
positive in spontaneously broken supersymmetry. 
In addition, the tree level
supertrace formulae \cite{Ferrara:1979wa}
made it difficult to consistently push up the
superpartners to a high mass scale and there was no known sensible
mechanism to generate gaugino masses. A new symmetry (R-symmetry) was
needed to assure proton stability, but the necessary breakdown of that
symmetry led to problems with unacceptable axions. It soon became clear
that realistic models had to be quite complicated and that supersymmetry
breakdown should be somewhat remote from the standard model itself.

With this in mind I decided 
that the right thing to do was to learn supergravity. After all gravity
existed in nature and should be included sooner or later in the
framework of the standard model. It was particularly imppressive
for me that
the local version of supersymmetry included gravity automatically (with
the gravitino as gauge particle). In fall 1981 I was moving to CERN as a
fellow and  this was the right place to learn supergravity. The
papers to read were \cite{Cremmer:1978iv,Cremmer:1979hn}
 and two of the authors, Sergio Ferarra and Luciano
Girardello were at CERN.

In this talk I would like to present the origin and status of hidden
sector supergravity in three steps. The suggestion of the scheme 
\cite{Nilles:1982ik}
came in
spring 1982 and led to hectic activities, so that in the beginning of
1983 the picture of what is today called ``gravity mediation'' was pretty
complete. A second important step of the development came in 1985 after
the ``string revolution'' initiated by Green and Schwarz \cite{Green:1984sg}. 
Here the ``hidden
$E_8$-sector'' of the heterotic string played a crucial role and the moduli
fields of the $d=10$ dimensional string theory entered the stage. This
will be discussed in chapter 3. More recently, with the advances of
string dualities and the formulation of M-theory 
\cite{Witten:1995ex,Hull:1995ys}
some of the remaining
questions could be clarified. The hidden sector now became a hidden wall
and this version of the scheme will be presented in chapter 4.

\section{1982: Hidden Sector Supergravity}

In a supersymmetric framework we wanted to understand the breakdown
scale of $SU(2)\times U(1)$ (thus the W- and Z-mass) as a consequence of
supersymmetry breakdown ($M_{\rm S}$). The hierarchy problem was rephrased:
why is $M_{\rm S}$ so small compared to the grand unified scale
 $M_{\rm GUT}$ or the
Planck scale $M_{\rm Planck}$? A satisfactory 
scheme would have been a dynamical of
breakdown supersymmetry, such as fermion-antifermion condensation 
or the condensation of gauge fermion 
bilinears: gaugino
condensation. Unfortunately, developments in 1981 led to the conclusion,
that in the simplest model, globally supersymmetric pure SU(N) gauge
theory, supersymmetry was not broken by the formation of gaugino condensates
\cite{Witten:1982df,Veneziano:1982ah,Nilles:1982my}.

Meanwhile my study of supergravity made progress and I could handle some
simple models. As an exercise I considered the effective
Langrangian for pure supersymmetric Yang-Mills theory as formulated by
Veneziano and Yankielowicz \cite{Veneziano:1982ah}
in its local supersymmetric version. Some
modifications had to be made to fine-tune the cosmological constant. The
result was surprising: supergravity was broken in the presence of a
gaugino condensate \cite{Nilles:1982ik}. 
Apart from the Polonyi superpotential 
\cite{Polonyi:1977pj} we had a
second example of local supersymmetry breakdown with a vanishing
cosmological constant.

\def\be{\begin{equation}}
\def\ee{\end{equation}}

But how could that be consistent with the apperent no-go theorem in the
globally supersymmetric theory? Let us assume that $\Lambda$, the
renormalization group invariant scale of the gauge interaction sets the
scale for gaugino condensation, thus
\be
<\chi\chi>  \approx \Lambda^3    
\ee
where $\chi$ denotes the gaugino. The supersymmetry breakdown scale 
$M_{\rm SUSY}$ (as
a vacuum expectation value of an auxiliary field) was here given by
\be
M_{\rm SUSY}^2 \approx {\Lambda^3 \over M_{\rm P}}    
\ee
where $M_{\rm P}$ denotes the Planck scale.

Thus everything was consistent, in the global limit 
$M_{\rm P} \rightarrow \infty$ 
one obtains $M_{\rm SUSY} \rightarrow 0$ and unbroken supersymmetry. In
spontaneously broken supergravity the gravitino becomes massive:
\be
m_{3/2} \approx {\Lambda^3 \over M_{\rm P}^2} 
\ee
This was very exciting since we had the possibility of dynamically
broken supergravity in a ``hidden sector'' with an additional Planck mass
suppression for the gravitino mass. What would that mean in case of
coupling to the ``observable sector'' (the supersymmetric extension of the
standard model). In the Polonyi case this had not been studied yet: in
any case one expected 
(at that time!) Str$M^2=0$ for the chiral multiplets and (in addition) no
convincing way to generate gaugino masses (i.e.  for the gauginos in the
observable sector).

In the present case, however, there was a way to generate gaugino masses
in the observable sector. Supergravity contains 4-fermion interactions
\be
 {h(\phi) \over M_{\rm P}^2} (\chi\chi)(\chi\chi), 
\ee
where $h(\phi)$ is a function of scalar fields $\phi$.

Imagine now a mixing term with one pair of observable sector gauginos
(like the gluinos, winos, etc.) and a second pair of hidden sector
gauginos with a nonvanishing vacuum expectation value. With this one obtains
soft supersymmetry breaking gaugino masses in the observable sector and
a completely satisfactory model \cite{Nilles:1982ik}. 
Masses for squarks and sleptons appear
via radiative corrections from the soft gaugino masses. The scale of
soft supersymmetry breaking parameters 
is determined by the gravitino mass
(with potential suppression factors). A detailed discussion can be found
in ref. \cite{Nilles:1983xx}. 
In this way the general picture of hidden sector
supergravity breakdown (or gravity mediated supersymmetry breakdown)
emerged. One needs two sectors: 

\begin{itemize}

\item the {\bf observable sector} including
quarks, leptons, gauge particles and Higgs bosons as well as their
supersymmetric partners and

\item  a {\bf hidden sector} coupled 
to the observable sector through interactions of
gravitational strength.

\end{itemize}

The hidden sector provides supersymmetry
breakdown at an intermediate scale and leads to a gravitino mass at a
rather low scale. Soft breaking terms of order of the gravitino mass (or
a few orders of magnitude smaller) appear in the observable sector and lead
to a phenomenological satisfactory version of the supersymmetric
extension of the standard model of strong and electroweak interactions.

Up to this point the supergravity action had only been constructed for
one chiral multiplet. A more detailed analysis of the low energy
effective action would require a generalization of this work. 
Such a generalization was  subsequently presented by
Cremmer, Ferrara, Girardello and van Proyen 
\cite{Cremmer:1982wb} stressing in particular
that the supertrace formulae of global supersymmetry was no longer valid in
this case. Chamseddine, Arnowit and Nath \cite{Chamseddine:1982jx}
quoted similar results in the
framework of a grand unified theory, while Barbieri, 
Ferrara and Savoy \cite{Barbieri:1982eh}
presented a supersymmetric extension of the standard model with soft
scalar mass terms, exploiting the new Str$M^2$ formulae in local
supersymmetry. Srednicki, Wyler and myself \cite{Nilles:1983dy}
completed this analysis of
soft SUSY breaking terms with among others the introduction of the
A-parameters. The relevance of the gauge kinetic function ($f$) for the
generation of soft gaugino masses was pointed out in 
ref. \cite{Ferrara:1983qs}. A general
comprehensive discussion was given in \cite{Hall:1983iz,Soni:1983rm}. 
So within nine months the full
picture had been completed. We now had a consistent scheme with
(dynamically) broken supersymmetry and a vanishing cosmological constant
but without tachyonic instabilities or harmful axions. Gaugino masses and
A-parameters were generated naturally: a consequence of the breakdown of
continuous  R-symmetry to discrete R-parity. Problems with flavour
changing neutral currents  (the so-called flavour problem) could be
avoided by assuming certain 
degeneracies of soft scalar mass terms, which
appeared automatically in case of gaugino masses as the dominant source
of supersymmetry breakdown.

Still today ``Hidden Sector Supergravity'' appears as the most promising
d=4 set-up for a low energy effective theory that extends the standard
model and avoids the hierarchy problem.

Remains the question of how this fits into ``Grand Scenario'', i.e.
Strings and/or extra dimensions.

\section{1984: Towards Higher Dimensions and String Theory}

In their classic paper \cite{Green:1984sg}
on anomaly cancellation in $S0(32)$
superstring theory, Green and Schwarz initiated a vast
research effort in that field. Very soon new string constructions,
such as the heterotic string theories \cite{Gross:1985fr,Gross:1986rr}
appeared and gave hope
for an application in particle physics. Especially the
heterotic $E_8\times E_8$ theory seemed very promising in
this respect. Compactification of six dimensions on a Ricci
flat manifold was shown to lead to supersymmetric models
in $d=4$ dimensions that could serve as candidates for a
supersymmetric extension of the standard model
\cite{Candelas:1985en}. In its
simplest form (the so-called standard embedding) this 
led to a theory with gauge group $E_6\times E_8$ and
quarks and lepton superfields in the 27-dimensional
representation to $E_6$. Thus all the observable fields
were contained in the $E_6$ sector.

With Jean Pierre Derendinger and Luis Ibanez at CERN, we were
excited about this fact. We had now a candidate theory which 

\begin{itemize}

\item[(i)] naturally contained an observable ($E_6$) sector
and a hidden ($E_8$) sector, that were

\item[(ii)] connected only via interactions of gravitational strength, and

\item[(ii)] in the simplest model the hidden sector
consisted of a pure supersymmetric gauge theory.

\end{itemize}

This was exactly the picture of hidden sector supergravity
we have discussed in the previous section: supersymmetry is 
potentially broken
by gaugino condensation in the hidden sector. It naturally
appeared in the framework of superstring theories. Still,
it was necessary to clarify the details. For this one needed a
simple way to arrive at an at least qualitative description  
of the low energy effective action. Such a method was supplied
in the scheme of reduction and truncation as defined in 
ref. \cite{Witten:1985xb}
which we will describe shortly. With this machinery at hand we 
could explicitely discuss the question of supersymmetry
breakdown in this set-up, as given in ref. \cite{Derendinger:1985kk}.
Similar results were
found independently in ref. \cite{Dine:1985rz}.
What we see is that the
messenger fields that are responsible for the mediation of
supersymmetry breakdown are a combination of the string theory
dilaton and moduli fields that describe the size and shape of
the extra compactified dimensions. In the simple case
considered here they are given by the 
moduli fields $S$ and $T$ that will be defined
below. We shall be quite brief in our discussion here and
refer the reader to ref. \cite{Derendinger:1986cv} for more details. 
For a more
recent review see ref. \cite{Nilles:1998uy}.

\def\NCA{\em Nuovo Cimento}
\def\NIM{\em Nucl. Instrum. Methods}
\def\NIMA{{\em Nucl. Instrum. Methods} A}
\def\NPB{{\em Nucl. Phys.} B}
\def\PLB{{\em Phys. Lett.}  B}
\def\PRL{\em Phys. Rev. Lett.}
\def\PRD{{\em Phys. Rev.} D}
\def\ZPC{{\em Z. Phys.} C}
% Some other macros used in the sample text
\def\st{\scriptstyle}
\def\sst{\scriptscriptstyle}
\def\mco{\multicolumn}
\def\epp{\epsilon^{\prime}}
\def\vep{\varepsilon}
\def\ra{\rightarrow}
\def\ppg{\pi^+\pi^-\gamma}
\def\vp{{\bf p}}
\def\ko{K^0}
\def\kb{\bar{K^0}}
\def\al{\alpha}
\def\ab{\bar{\alpha}}
\def\be{\begin{equation}}
\def\ee{\end{equation}}
\def\bea{\begin{eqnarray}}
\def\eea{\end{eqnarray}}
\def\CPbar{\hbox{{\rm CP}\hskip-1.80em{/}}}%temp replacement due to no font

\def\be{\begin{equation}}
\def\ee{\end{equation}}
\def\ba{\begin{array}}
\def\ea{\end{array}}
\def\bea{\begin{eqnarray}}
\def\eea{\end{eqnarray}}
\def\GeV{{\rm GeV}}
\def\tr{{\rm tr}}
\def\thefootnote{\fnsymbol{footnote}}
\def\chib{{\bar\chi}}
\def\psib{{\bar\psi}}
\def\nn{\nonumber}

\def\wS{S}
\def\wT{T}                                                  .

\def\sS{{\cal S}}
\def\sT{{\cal T}}

\def\NPB#1#2#3{{Nucl.~Phys.} {\bf{B#1}} (19#2) #3}
\def\PLB#1#2#3{{Phys.~Lett.} {\bf{B#1}} (19#2) #3}
\def\PRD#1#2#3{{Phys.~Rev.} {\bf{D#1}} (19#2) #3}

\subsection{Supersymmetry breakdown in heterotic string theory}

In our review of supersymmetry breakdown in the $d=10$ weakly coupled 
$E_8 \times E_8$ theory we shall start from the $d=10$ effective field 
theory and go to $d=4$ dimensions via the method of reduction and truncation 
explained in ref. \cite{Witten:1985xb}. In string theory compactified on an 
orbifold this would describe the dynamics of the untwisted sector. 
We retain the usual moduli fields $\wS$ (the dilaton) and $\wT$ 
(the overal size of compactified extra dimensions) as well as matter 
fields $C_i$ that transform nontrivially under the observable sector 
gauge group. In this approximation, the K\"ahler potential is given by 
\cite{Witten:1985xb,Derendinger:1986cv}
\be
G = - \log (\wS + \bar \wS) - 3 \log (\wT+\bar \wT -2C_i\bar C_i) + 
\log \left| W \right|^2  
\label{eq:G}  
\ee
with superpotential
\be
 W(C) = d_{ijk} C_iC_jC_k   
\ee
and the gauge kinetic function is given by the dilaton field
\be
f = \wS \,. 
\ee
We assume  the formation of a gaugino condensate
$<\chi\chi>  = \Lambda^3$ 
in the hidden sector
where $\Lambda$ is the renormalization group invariant scale of the 
confining hidden sector gauge group. The gaugino condensate appears 
in the expression for the auxiliary components of the chiral superfields 
\cite{Ferrara:1983qs}
\be
F_j = \left( G^{-1} \right)_j^k 
      \left( \exp (G/2) G_k +{1\over 4} f_k (\chi\chi) \right) + \ldots 
\ee
which are order parameters for supersymmetry breakdown. Minimizing the 
scalar potential we find $F_\wS = 0$, $F_\wT \ne 0$ and a vanishing 
cosmological constant. Supersymmetry is broken and the gravitino mass 
is given by  \cite{Nilles:1990zd}
\be
m_{3/2} = \frac{\left< F_T \right>}{\wT + {\bar \wT}} 
\approx {\Lambda^3 \over M_P^2} 
\label{eq:m32}
\ee
and $\Lambda = 10^{13}$ GeV would lead to a gravitino mass in the 
TeV -- range. A first inspection of the soft breaking terms in the 
observable sector gives a surprising result. They vanish in this 
approximation. Scalar masses are zero because of the no--scale structure
\cite{Ellis:1984ei} 
in (\ref{eq:G}) (coming from the fact that we have only included fields 
of modular weight $-1$  under T--duality in this case 
\cite{Brignole:1994dj}). 
In a more general situation we might expect scalar masses $m_0$ comparable to 
the gravitino mass $m_{\rm 3/2}$ and the above result  $m_0=0$ 
would just be an 
artifact of the chosen approximation at the classical level. 
Thus soft scalar masses are strongly model dependent. 
Gaugino masses $m_{1/2}$  are easier to discuss. They are given by
\be
m_{1/2} = \frac{ {\partial f \over \partial \wS} F_\wS 
                +{\partial f \over \partial \wT} F_\wT }
               {2 {\rm Re} f}                          
\label{eq:m12}
\ee
and with $f = \wS$ and $F_\wS = 0$ we obtain $m_{1/2} = 0$. One loop 
corrections will change this picture as can be seen already by an 
inspection of the Green--Schwarz anomaly cancellation counter terms, 
as they modify $f$ at one loop. In the simple example of the so--called
standard embedding with gauge group $E_6 \times E_8$ we obtain 
\cite{Derendinger:1986cv,Ibanez:1986xy}
\be
f_6 = \wS +\epsilon \wT\,;
\qquad\qquad 
f_8 = \wS - \epsilon \wT\,.
\label{eq:f6f8}
\ee
This dependence of $f$  on $\wT$ will via (\ref{eq:m12}) lead to 
nonvanishing gaugino masses which, however, might be small compared 
to $m_{3/2}$ and $m_0$ since $\epsilon \wT$ is considered 
a small correction to the classical result. This might be   
problematic when applied to the class of supersymmetric extension of 
the standard model, where $m_0$ is
large compared to $m_{1/2}$. With the large difference of the soft
scalar and the gaugino masses a sizeable fine tuning
is needed to induce the breakdown of electroweak symmetry
at the correct scale \cite{Nilles:1984ge}.
The smallness of the gaugino masses might also lead to a problem in
the context of relic abundances of the lightest
superparticles (LSP) \cite{Jungman:1996df}. Ideally one would like to
consider again a model where the main source of supersymmetry breaking
is given by the gaugino masses.

\subsection{The role of the antisymmetric tensor field}

There are a few things that should be retained from this discussion 
that are not so transparent in the short presentation given above.
In particular it is the role of the two index antisymmetric
tensor field $B_{MN}$ that is important. In $d=10$ supergravity
the (3 index) field strength $H_{MNP}$ is not just the curl of $B$ but has
to be supplied with so-called Chern Simons terms
\cite{Chamseddine:1981ez}.

We now want to stress the following points that will also be
important in the discussion in the next section.

\begin{itemize}

\item The superpotential of the $d=4$ effective action
originates in the co-called Yang-Mills-Chern-Simons term $\omega^{\rm YM}$.  

\item The Green Schwarz counterterms of anomaly
cancellation, will provide (among others) a one-loop corection to
the gauge kinetic function, that is classicaly just given by $S$

\item In $d=4$ the real part of $S$ is given by $g^{-2}$, thus the gauge
coupling constant $g$ is a field dependent quantity.

\item In the $d=10$ dimensional theory the field strength $H$ and the
gaugino bilinears $(\chi\chi)$
appear in a complete square, such that at the classical 
level the potential is positive definite.

\item As a result of this a nonvanishing vacuum
expectation value of $H$ will trigger 
supersymmetry breakdown and the dilaton will adjust its
vacuum expectation value such that the cosmological
constant vanishes.

\item A vanishing value of $<H>$ will then lead to a situation
where $S$ runs to infinity with a vanishing gauge coupling constant.

\end{itemize}

\subsection{Open questions}

Certainly the potential runaway of the dilaton $S$ is an open
problem in this scenario (as in any theory based
on perturbative string theory ), and up to today no really satisfactory
solution has been proposed. Other problems concern the exact form of the
scalar potential. While at the classical level the vacuum energy vanishes,
this is certainly not obvious in the quantum theory. As a result of this,
it is quite difficult to obtain a statement about even the
qualitative form of the potential in a model independent way.
This, of course, implies that the soft supersymmetric breaking parameters
(like squark and slepton masses as well as A-parameters)
are strongly model dependent. In contrast, the situation for the
gaugino masses is much simpler (as usual) and maybe they could even
be dominant in a realistic framework. Their origin in the simple
set-up is a direct consequence of the Green-Schwarz counterterms
in the low-energy effective action. In perturbative string theory
they can be obtained through a one loop calculation 
\cite{Dixon:1991pc} that agrees
with the former result in the limit of large $T$.

In a next step, one would like to use more explicit input from
string theory. One approach would be to exploit the presence of
stringy symmetries \cite{Nilles:1986cy,Lauer:1989ax}, 
like $T$-duality, that should be valid in
perturbative string theory. In fact, this picture of $T$-duality
can be described successfully for the classical effective action
\cite{Font:1990nt,Nilles:1990jv,Ferrara:1990ei,Binetruy:1991ck}.
There appear, however, problems when one incorporates loop effets,
as now the $S$ and $T$-moduli mix. A successful implemenation of
$T$-duality then seems to require an understanding of $S$-duality
as well. $S$-duality, however, is not an explicit symmetry of
the perturbative low energy effective action, and this is 
probably the reason, why this approach can not be completed
\cite{Lalak:1995hn,Lalak:1995zj,Lalak:1999bk}.
The problem of the runaway dilaton, and thus the determination
of the vacucum expextation value of $S$ seems to require a 
nonperturbative treatment and no really convincing scheme has 
emerged. As we have seen above this problem might be intimately
related to the origin of a vacuum expectation value of the
three-index tensor field strengh $<H>$ and the apearence of 
Chern-simons terms. Still, a better understanding of string
symmetries would be the road to a better control of
the low energy effective action.

\section{1995: Towards $d=11$ and (Heterotic) M--theory}

A new step towards the understanding of nonperturbative dualities
in string theory came in 1995 with the formulation of the
concept of M-theory \cite{Witten:1995ex,Hull:1995ys}
and its embedding in $d=11$ dimensional
space time. For our purposes we are especially interested in the
$E_8\times E_8$ version as constructed by Horava and Witten
\cite{Horava:1996qa,Horava:1996ma}.
The new aspect of this picture is the fact that now hidden and
observable sector become separated geometrically. Thus
hidden sector supergravity breakdown now becomes the breakdown
of supergravity at a hidden brane, and again gaugino condensation
is the preferred mechanism as suggested by
Horava \cite{Horava:1996vs} and 
Olechowski, Yamaguchi and myself \cite{Nilles:1997cm}. 
Early related work has been done in \cite{Lalak:1998zu,Lukas:1998fg}

Let us now reconsider these
questions in the  strongly coupled $E_8 \times E_8$ -- $M$--theory
 in somewhat more detail. 
The effective action is given by 
(for details see \cite{Horava:1996ma})
\bea
L =
{1\over \kappa^2} \int
d^{11}x \sqrt{g}
\left[
       - \frac{1}{2}R +\ldots 
\right] + \ldots \qquad
\\
+\frac{1}{2\pi(4\pi\kappa^2)^{2/3}}  \int 
d^{10}x \sqrt{g}
\left[ 
    - \frac{1}{4} F^a_{AB} F^{aAB} +\ldots
\right].
\nn
\eea
Compactifying to $d=4$ we obtain 
\cite{Witten:1996mz} (with the correction pointed 
out in ref. \cite{Conrad:1998ww})
\bea
G_N &=&   8\pi \kappa_4^2 = {\kappa^2 \over 8\pi^2 V \rho}\,,
\\
\alpha_{GUT} &=& {(4\pi\kappa^2)^{2/3} \over V}
\label{eq:GN}
\eea
with $V= R^6$ and $\pi\rho= R_{11}$. Fitting $G_N$ and $\alpha_{GUT}=1/25$ 
then gives $R_{11} M_{11}$ of order 10 and 
$M_{11} R\approx 2.3$.
The rather large value of the $d=4$ reduced Planck Mass  
$M_{P}= \kappa_4^{-1}$ is obtained as a result of the fact that $R_{11}$
is large compared to $R$.

We now perform a compactification using the method of reduction and 
truncation as above. For the metric we write 
\cite{Nilles:1997cm,Nilles:1998sx}
\be
g_{MN} = 
\left(
\ba{ccc}
e^{-\gamma} e^{-2\sigma} g_{\mu\nu} & & \\
 & e^\sigma \delta_{mn} & \\
 & & e^{2\gamma} e^{-2\sigma} 
\ea
\right)
\ee
with 
$M,N = 1 \ldots 11$; $\mu,\nu = 1 \ldots 4$; $m,n = 5 \ldots 10$; 
$2R_{11} = 2\pi\rho  = M_{11}^{-1} e^\gamma e^{-\sigma}$  and
$V= e^{3\sigma} M_{11}^6$.  At the classical level this leads to a 
K\"ahler potential as in (\ref{eq:G}) 
\be
K = - \log (\sS + {\bar \sS}) - 3 \log (\sT+{\bar \sT} -2C_i {\bar C_i})
\ee
with
\bea
\sS &=& \frac{2}{\left( 4\pi \right)^{2/3}} 
\left( e^{3\sigma} \pm i 24 \sqrt{2} D \right)
\,,
\\
\sT &=& \frac{\pi^2}{\left( 4\pi \right)^{4/3}}
\left( e^{\gamma} \pm i 6 \sqrt{2} C_{11} \right)
\eea
where $D$ and $C_{11}$ fields are defined by
\bea
\frac{1}{4!} e^{6\sigma} G_{11\lambda\mu\nu} &=& 
\epsilon_{\lambda\mu\nu\rho}\left(\partial^\rho D \right)
\,,
\\
C_{11 i {\bar j}} &=& C_{11} \delta_{i \bar j}
\eea
and $x^i$ ($x^{\bar j}$) is the holomorphic (antiholomorphic) coordinate 
of the Calabi--Yau manifold. The imaginary part of $\sS$ (Im$\sS$) 
corresponds to the model independent axion, and the gauge kinetic 
function is $f = \sS$. This is very similar to the weakly coupled case. 
Before drawing any conclusion from these formulae, however, we have to 
discuss a possible obstruction at the one loop level. It can be understood 
from the mechanism of anomaly cancellation 
\cite{Witten:1996mz}. 
For the 3--index tensor field $H$ in $d=10$ supergravity to be well defined 
one has to satisfy $dH = \tr F_1^2 + \tr F_2^2 - \tr R^2 = 0$ 
cohomologically. In the simplest case of the standard embedding one 
assumes $\tr F_1^2 = \tr R^2$ locally and the gauge group is broken to 
$E_6 \times E_8$. Since in the M--theory case the two different gauge 
groups live on the two different boundaries of space--time such a 
cancellation point by point is no longer possible. We expect nontrivial 
vacuum expectation values (vevs) of 
\be
(dG) \propto \sum_i \delta(x^{11} - x^{11}_i) 
\left( \tr F_i^2 - {1\over 2} \tr R^2 \right)
\ee
at least on one boundary ($x^{11}_i$ is the position of $i$--th boundary). 
In the case of the standard embedding we would have 
$\tr F_1^2 - {1\over 2} \tr R^2 = {1\over 2} \tr R^2$ on one and 
$\tr F^2_2 - {1\over 2} \tr R^2 = - {1\over 2} \tr R^2$ on the other 
boundary. This might pose a severe problem since a nontrivial vev  of 
$G$ might be in conflict with supersymmetry ($G_{11ABC}=H_{ABC}$). 
The supersymmetry 
transformation law in $d=11$ reads  
\be
\delta \psi_M 
=
D_M\eta + \frac{\sqrt{2}}{288} G_{IJKL} 
          \left( \Gamma_M^{IJKL} +\ldots \right) \eta
+ \ldots
\label{eq:dpsiM}
\ee
Supersymmetry will be broken unless e.g.\ the derivative term $D_M\eta$ 
compensates the nontrivial vev of $G$. Witten has shown 
\cite{Witten:1996mz} 
that such a cancellation can occur and constructed the solution in the 
linearized approximation (linear in the expansion parameter $\kappa^{2/3}$) 
which corresponds to the large $T$--limit in the weakly coupled 
theory\footnote
{For a discussion beyond this approximation in the weakly coupled 
case see refs.
\cite{Nilles:1997vk,Stieberger:1999yi}.}.

The supersymmetric solution 
can be shown to lead 
\cite{Nilles:1997cm} to a nontrivial dependence of the 
$\sigma$ and $\gamma$ fields with respect to $x^{11}$: 
\be
{{\partial\gamma}\over{\partial x^{11}}} 
=
- {{\partial\sigma}\over{\partial x^{11}}} 
=
\frac{\sqrt{2}}{24}
\frac
{\int d^6x \sqrt{g} \omega^{AB}\omega^{CD}G_{ABCD}}
{\int d^6x \sqrt{g}}
\label{eq:19}
\ee
where the integrals are over the Calabi--Yau manifold and $\omega$ is 
the corresponding K\"ahler form. Formula (\ref{eq:19}) contains
all the information to deduce the effective 
$d=4$ supergravity theory, including the
K\"ahler potential and the gauge kinetic function
\cite{Nilles:1997cm,Nilles:1998sx}.
A definition of our $\sS$ and $\sT$ 
fields in the four--dimensional theory now requires an average 
over the 11--dimensional interval. We  therefore write
\bea
\sS &=& \frac{2}{\left( 4\pi \right)^{2/3}} 
\left( e^{3\bar\sigma} \pm i 24 \sqrt{2} \bar D \right)
\,,
\\
\sT &=& \frac{\pi^2}{\left( 4\pi \right)^{4/3}}
\left( e^{\bar\gamma} \pm i 6 \sqrt{2} \bar C_{11} \right)
\label{eq:sT}
\eea
where bars denote averaging over the 11th dimension. It might be of some 
interest to note that the combination $\sS\sT^3$ is independent of 
$x^{11}$ even before this averaging procedure took place.

$\exp(3\sigma)$ represents the  volume of the six--dimensional compact 
space in units of $M_{11}^{-6}$. The $x^{11}$ dependence of $\sigma$ 
then leads to the geometrical picture that the volume of this space 
varies with $x^{11}$ and differs at the two boundaries. In the given 
approximation, this variation is linear, and for growing $R_{11}$ the 
volume on the $E_8$ side becomes smaller and smaller. At a critical 
value of $R_{11}$ the volume will thus vanish and this will provide 
us with an upper limit on $R_{11}$. For the phenomenological 
applications we then have to check whether our preferred choice of $R_{11}$
that fits the correct value of the $d=4$ Planck 
mass\footnote
{With $V$ depending on $x^{11}$ we have to specify which values 
should be used in eq.\ (\ref{eq:GN}). The appropriate choice 
in the expression for $G_N$ is the average value of $V$ 
while in the expression for $\alpha_{GUT}$
we have to use $V$ evaluated at the $E_6$ border.} 
satisfies this bound. Although the coefficients are model dependent we 
find in general that the bound can be satisfied, but that $R_{11}$ is 
quite close to its critical value. A choice of $R_{11}$ much larger than 
$({\rm few}\times 10^{15} \GeV)^{-1}$ is therefore not permitted.

This variation of the volume is the analogue of the one loop correction 
of the gauge kinetic function (\ref{eq:f6f8}) in the weakly coupled case 
and has the same origin, namely a Green--Schwarz anomaly cancellation 
counterterm. In fact, also in the strongly coupled case we find 
\cite{Nilles:1997cm}, 
as a consequence of (\ref{eq:19}), corrections 
for the gauge coupling constants at the $E_6$ and $E_8$ side.

Gauge couplings will no longer be given by the (averaged) $\sS$--field, 
but by that combination of the (averaged) $\sS$ and $\sT$ fields which 
corresponds to the $\sS$--field before averaging at the given boundary: 
\be  
f_{6,8} = \sS \pm \alpha \sT
\ee
at the $E_6$ ($E_8$) side 
respectively\footnote
{With the normalization of the $\sT$ field as in (\ref{eq:sT}), 
$\alpha$ is a quantity of order unity.}.
The critical value of $R_{11}$ will correspond to infinitely strong 
coupling at the $E_8$ side $\sS - \alpha \sT = 0$ (Notice the similarity 
to (\ref{eq:f6f8}) in the weakly coupled limit). Since we are here close 
to criticality a correct phenomenological  fit of 
$\alpha_{\rm GUT} = 1/25$ should include this correction 
$\alpha_{\rm GUT}^{-1} = \sS + \alpha \sT$ where $\sS$ and 
$\alpha \sT$ give comparable contributions. This is a difference to the 
weakly coupled case, where in $f= \wS + \epsilon \wT$ the latter 
contribution was small compared to $\wS$. Observe that this picture of 
a loop correction $\alpha \sT$ to be comparable to the tree level result 
still makes sense in the perturbative expansion, since $f$ does not 
receive further perturbative corrections beyond one loop 
\cite{Nilles:1986cy}.
 
\subsection{Supersymmetry breakdown in M--theory}

In a next step we are now ready to discuss the dynamical breakdown 
of supersymmetry via gaugino condensation in the strongly coupled 
M--theory picture. In analogy to the previous discussion we start 
investigating supersymmetry transformation laws in the higher--dimensional 
(now $d=11$) field theory 
\cite{Horava:1996vs}:
\bea
\delta \psi_A
&\sim &
D_A\eta 
+ G_{IJKL} 
  \left( \Gamma_A^{IJKL} - 8 \delta_A^I \Gamma^{JKL} \right) \eta
\nn\\
&-& 
 \delta(x^{11})
  \left( \chib^a \Gamma_{BCD} \chi^a \right) 
  \Gamma_A^{BCD} \eta
  + \ldots
\\
\delta \psi_{11}
&\sim&
D_{11} \eta +  G_{IJKL}
\left( \Gamma_{11}^{IJKL} - 8 \delta_{11}^I \Gamma^{JKL} \right) \eta
\nn\\
&+&
  \delta(x^{11}) \left( \chib^a \Gamma_{ABC} \chi^a \right) \Gamma^{ABC} \eta
  + \ldots
\eea
where gaugino bilinears appear in the right hand side of both expressions. 
It can therefore be expected that gaugino condensation breaks supersymmetry. 
Still the details have to be worked out. In  the $d=10$ example, the 
gaugino condensate and the three--index tensor field $H$ contributed to 
the scalar potential in a full square. This has led to a vanishing 
cosmological constant as well as the fact that $F_\wS =0$ at the classical 
level. Ho\v{r}ava has observed 
\cite{Horava:1996vs}
that a similar mechanism might be in operation in the $d=11$ theory
After a careful calculation this leads to a vanishing variation 
$\delta\psi_A=0$. In our model (based on reduction and truncation) 
we can now compute these quantities explicitly. We assume gaugino 
condensation to occur at the $E_8$ boundary 
\be
\left<\chib^a \Gamma_{ijk} \chi^a \right> = \Lambda^3 \epsilon_{ijk}
\ee
where $\Lambda < M_{\rm GUT}$ and $\epsilon_{ijk}$ is the covariantly 
constant holomorphic 3--form. This leads to a nontrivial vev of 
$G_{11ABC}$ at this boundary and supersymmetry is 
broken\footnote
{One might speculate that a nontrivial vev of $D_A\eta$ might be operative 
here as in the case without gaugino condensation (see discussion after 
eq.\ (\ref{eq:dpsiM})). However, the special values of 
$H_{ijk} \propto \epsilon_{ijk}$ necessary to cancel the contribution of the 
gaugino condensate do not permit such a mechanism (see footnote 6 in ref.\ 
\cite{Witten:1996mz}).}. 
At that boundary we obtain $F_\sS=0$ and $F_\sT \ne 0$ as expected from 
the fact that the component $\psi_{11}$ of the 11--dimensional gravitino 
plays the role of the goldstino.

In the effective $d=4$ theory we now have to average over the 11th 
dimension leading to
\be
\left< F_\sT \right> 
\approx
\frac{1}{2} \sT \frac{\int dx^{11} \delta\psi_{11}}{\int dx^{11}}
\ee
as the source of SUSY breakdown.

This will then allow us to compute the 
size of supersymmetry breakdown on the observable $E_6$ side. Gravitational 
interactions play the role of messengers that communicate between the two 
boundaries. This effect can be seen from (\ref{eq:GN}): large $R_{11}$ 
corresponds to large $M_P$ and $\left< F_\sT \right>$ gives the effective 
size of SUSY breaking on the $E_6$ side ($R_{11} \rightarrow\infty$ implies 
$M_P \rightarrow\infty$). The gravitino mass is given by 
\be
m_{3/2} 
= {\left< F_\sT \right> \over \sT + \bar \sT} 
\approx {\Lambda^3 \over M_P^2}
\ee
(similar to (\ref{eq:m32}) in the weakly coupled case) and we expect this 
to represent the scale of soft supersymmetry breaking parameters in the 
observable sector \cite{Nilles:1984ge}. 
These soft masses are determined  by the coupling of 
the corresponding fields to the goldstino multiplet. As we have seen before, 
we cannot compute the scalar masses reliably. In 
the naive approximation used here we obtain 
$m_0=0$ because of the no--scale structure \cite{Ellis:1984ei}.
This might be an artifact 
of the approximation. Fields of different modular weight will receive 
contribution to $m_0$ of order $m_{3/2}$. For the mass of a field 
$C$ we have  
\cite{Brignole:1994dj}
\be
m_0^2 = m_{3/2}^2 - F^i {\bar F}^{\bar j} 
\frac{Z_{i \bar j} - Z_i Z^{-1} {\bar Z_{\bar j}}}{Z}
\ee
where $i,j = \sS, \sT$ and $Z$ is the moduli dependent coefficient of 
$C \bar C$ term appearing in the K\"ahler potential. Scalars of 
modular weight $-1$ will become massive through radiative corrections. 
This then leads to the expectation that $m_{3/2}$ should be in the 
TeV--region and $\Lambda\approx 10^{13}$ GeV\ \footnote
{In realistic models $E_8$ is broken and $\Lambda$ is adjusted by model 
building.}. 
So far this is all similar to the weakly coupled case.

\subsection{Gaugino masses and WIMPs}

An important difference appears, however, when we turn to the discussion 
of observable sector gaugino masses (\ref{eq:m12}). In the weakly coupled 
case they were zero at tree level and appeared only because of the 
radiative corrections at one loop (\ref{eq:f6f8}). As a result of this small 
correction, gaugino masses were expected to be much smaller than $m_{3/2}$. 
In the strongly coupled case the analog of (\ref{eq:m12}) is still valid 
\be
m_{1/2} = \frac{ {\partial f_6 \over \partial \sS} F_\sS 
                +{\partial f_6 \over \partial \sT} F_\sT }
               {2 {\rm Re} f_6}                          
\ee
and the 1--loop effect is encoded in the variation of the $\sigma$ and 
$\gamma$ fields from one boundary to the other. Here, however, the loop 
corrections are sizable compared to the classical result because of the 
fact that $R_{11}$ is close to its critical value. As a result we expect 
observable gaugino masses of the order of the gravitino mass. The problem 
of the small gaugino masses does therefore not occur in this situation. 
Independent of the question whether $F_\sS$ or $F_\sT$ are the dominant 
sources of supersymmetry breakdown, the gauginos will be heavy of the
order of the gravitino mass. Thus it could very well be that
gaugino masses are the dominant sources of supersymmetry breakdown
in the observable sector.
The exact relation between the soft breaking 
parameters $m_0$ and $m_{1/2}$ will be a question of model building. If 
in some models $m_0 \ll m_{1/2}$ this might give a solution to the flavor 
problem. The no--scale structure found above might be a reason for such 
a suppression of $m_0$. As we have discussed above, this structure, however, 
is an artifact of our simplified approximation and does not 
necessarily survive in 
perturbation theory. At best it could be kept exact (but only for the 
fields with modular weight $-1$) in the $R_{11} \rightarrow \infty$ limit. 
The upper bound on $R_{11}$ precludes such a situation. With observable 
gaugino masses of order $m_{3/2}$ we also see that $m_{3/2}$ cannot be 
arbitrarily large and should stay in the TeV -- range
\cite{Nilles:1997cm,Lalak:1998zu}.

The question of the soft gaugino masses is, of course, very important
for the determination of the mass of the lightest supersymmetric
particle (LSP). This might be stable and provide a candidate for the
dark matter found in the universe
(a so-called WIMP, for weakly interacting massive
particle). Given the present situation of the supersymmetric extension 
of the standard model, the LSP is most likely a gaugino of the
wino- or bino-type. The relic abundance of these neutralinos in the early
universe depends crucially on its mass and its coupling to ordinary
matter. The mass is predominantly determined through the soft
gaugino masses, while the interactions are mediated by the scalar
partners of leptons. Too high a mass  of the scalar partners
leads then to a weak coupling of the LSP to ordinary matter,
reducing the annihilation of WIMPs in the early universe; with the
potential danger of an overcritical WIMP density. For this relic density
of WIMPs to be both acceptable and cosmologically interesting, only
certain ranges of soft scalar masses and gaugino masses are 
allowed. It turns out, that the heterotic M-theory could 
lead to such an acceptable scenario (in contrast to the the
situation in the weakly coupled case). A detailed discussion can be
found in ref. \cite{Kawamura:1998fv}.

\subsection{The Super--Higgs mechanism}

In the previous sections the estimate of the gravitino mass 
was obtained \cite{Nilles:1997cm}
using a the simplified approximation according
to which the higher dimensional bulk fields were integrated out
via an averaging proceedure\footnote{A corresponding analysis in global
supersymmetry has been performed in ref. \cite{Mirabelli:1998aj}.}. 
In this picture, the goldstino
mode was represented by the lowest Kaluza--Klein $\Psi_0$ mode of a
higher dimensional field $\Psi$. In the super--Higgs mechanism this
mode supplies the additional degrees of freedom to render the
gravitino massive. Qualitatively this simplified approximation
does give a consistent picture, but there remain some open
questions and potential problems when one looks into details of
the super--Higgs mechanism. 
In this section we would like to point out these
potential problems and show how they can be resolved. 
For a detailed explanation we refer the reader 
to \cite{Meissner:1999ja}. We shall concentrate here
on a qualitative discussion of the mechanism involved.
 This will include
a discussion of the possible nature of the goldstino (is it a bulk
or a wall field), the relation to the Scherk-Schwarz mechanism 
\cite{Scherk:1979zr}
in that context \cite{Antoniadis:1997ic} and an upper limit for
the gravitino mass in the present picture. We shall argue that
a meaningful realization of the super--Higgs mechanism seems to 
require some modes in the bulk other than the graviton and the 
gravitino. Finally we shall comment on the phenomenological
consequences of this findings, including 
a discussion of the nature of the soft
breaking terms on both walls.

Specifically we want to address the following two questions:

\begin{itemize}

\item[(i)] the nature of the massless gravitino in the presence
of several $F-$terms\footnote{We generically use
the notation $F-$term for the source of supersymmetry breakdown.
Depending on the specific situation this could represent a
$D-$term or a gaugino condensate as well.}
 on different walls that cancel and lead to
unbroken supersymmetry,

\item[(ii)] the identification of the goldstino in the case of
broken supersymmetry.

\end{itemize}

The first question (i) arises because of a particalur nonlocal effect
of supersymmetry breakdown first observed by Horava \cite{Horava:1996vs}.
A given source of supersymmetry breakdown (parametrized by a
vacuum expectation value (vev) of an auxiliary field $F$) on one wall
could be compensated by a similar but opposite value $(-F)$ on 
another (separated) wall. Any calculation and approximation of
the system thus has to reproduce this behaviour. The
previously mentioned averaging proceedure over the bulk distance
does this in a trivial way, leading to unbroken 
supersymmetry as expected.
A detailed inspection of the gravitino, however, reveals a problem.
If we start with the situation $F=0$ it is easy to define the
massless gravitino $\Psi_0$ in the $d=4$ theory. Switching on a nontrivial
$F$ on one brane and $(-F)$ on the other still should give a
massless gravitino, but $\Psi_0$ turns out to be no longer a mass eigenstate.
The resolution of this problem and the correct identification of the
gravitino can be found in \cite{Meissner:1999ja}. It is a particular
combination of the possible gravitini that appear when one, for
example, reduces a 5-dimensional theory to a theory in $d=4$ on a
finite $d=5$ interval. The theory on a $d=5$ circle would lead to
$N=2$ supersymmetry in $d=4$ and two massless gravitini (zero modes on
the circle). The
$Z_2$ projection on the interval removes one of the gravitini
and is $N=1$ supersymmetric. A nonvanishing vev of $F$ now interferes
with the boundary conditions and the massless gravitino will be
a linear combination of the zero mode and all the excited 
Kaluza-Klein modes whose
coefficients will depend on $F$ (assuming, of course, unbroken
supersymmetry due to a compensating vev $-F$ on another wall).

The second question (ii) deals with the nature of the
goldstino (i.e. the longitudinal components of the gravitino)
in the case of broken supersymmetry. Remember that the
simplified averaging proceedure leads to a goldstino that
corresponds to the lowest Kaluza--Klein mode $\Psi_0$
of a higher-dimensional bulk field $\Psi$. Inspecting the
gravitino mass matrix in this case reveals the fact that this
field $\Psi_0$ is not a mass eigenstate, but mixes with
infinitely many higher Kaluza--Klein modes $\Psi_n$. A consistent 
manifestation of a super--Higgs mechanism would require a
diagonalization of this mass matrix and an identification of
the goldstino. This problem has been solved in \cite{Meissner:1999ja},
by a suitable redefinition of the Kaluza-Klein modes. 
Thus it is shown that a consistent Super-Higgs mechanism is
at work.

\section{The nature of hidden wall supersymmetry breakdown}

The resolution of these puzzles clarifies some of the other
questions of the approach. It also allows us to generalize
this to string theories of type I with D-branes and
supersymmetry breakdown on distant branes. The qualitative
picture, of course, is very similar. 

\begin{itemize}

\item  The nonlocality of the breakdown shows some resemblance
to the breakdown of supersymmetry via the 
Scherk--Schwarz \cite{Scherk:1979zr} mechanism. Here,
however, the real goldstino of the spontaneous breakdown of
supersymmetry can be unambiguously identified.

\item The possibility to cancel the supersymmetry breakdown
on a distant wall by a vev on the local wall tells us,
that the mass splittings of broken supersymmetry have to
be of order of the gravitino mass $m_{3/2}$ on {\it both}
walls.

\item In terms of the physical quantities there is no real
extra suppression, once we separate the walls by a large
distance $R$. In the limit $R\rightarrow\infty$
we will have $M_{\rm Planck}\rightarrow\infty$ as well.
The suppression of the soft breaking parameters will
always be gravitational.

\item In general, when we have a system of many separated 
branes with potential sources of supersymmetry breakdown,
the actual breakdown will be obtained by the sum of these
contributions. The averaging proceedure will be very
useful to decide whether supersymmetry is broken or not. The 
identification of the goldstino, however, is more
difficult and requires a careful calculation.

\item A successful implementation of the super--Higgs mechanism
will require some fields other than gravitino and graviton in 
the bulk\footnote{Usually they arise as modes of the higher
dimensional supergravity multiplet.}. 
This implies that in the absence of such fields
a consistent
spontaneous breakdown of supergravity might not be achieved.

\end{itemize}

We thus have a consistent scenario of supersymmetry breakdown
in the framework of heterotic M-theory. It provides us with explicit
formulae of soft supersymmetry breaking terms in the
low energy effective theory. Interestingly enough these 
formulae differ from those obtained in the weakly coupled 
heterotic $E_8\times E_8$ theory, most notably with respect
to the size of soft gaugino masses. The phenomenological
properties of the scenario, therefore have to be reconsidered
in various circumstances. This might include a discussion of
the mass and nature of the LSP, the question of the universality
of the soft scalar masses as well as possible
new solutions to the strong CP problem. More work needs to be done
to work out these questions in detail.
In summary we can say that
the picture of supersymmetry breakdown in the M--theoretic limit looks 
very promising. It is very similar to the weakly coupled case, 
but avoids the problem of the small gaugino masses. This has important
consequences for the phenomenological and cosmological properties
of the effective models in four dimensions.

\section{Outlook}

Since its original formulation in 1982, gravity mediation appeared
as a very attractive scheme to introduce spontaneous (dynamical)
supersymmetry breakdown into the supersymmetric extension of the
$SU(3)\times SU(2)\times U(1)$ standard model of strong, weak and
electromagnetic interactions. Its natural implementation in
(heterotic) string theory and M-theory stengthens the confidence
in this mechanism of hidden sector supergravity breakdown.
It gives us useful hints about the possibility of a local cancellation
of the classical vacuum energy, the role of anomaly cancellation
via the Green-Schwarz counterterms, the importance of underlying string
symmetries and the specific role of the antisymmetric tensor fields in
string- and M-theory.

Still there remain some open questions about the possible 
incorporation of $T$- and $S$-duality, the microscopic origin 
of the vacucum expectation value of the $<H>$ and $<G>$ fields and their
potential to avoid the run-away problem of the dilaton. The control
of the cosmological constant beyond the classical level remains
a problem and this makes it difficult to determine
the soft scalar mass terms in a model independent way. Soft gaugino masses
are easier to handle since their size seems to be controlled by
the mechanism of anomaly cancellation. The most attractive scheme
emerges in the case where gaugino masses are the dominant
source of supersymmetry breakdown.

%\ \\
%\newpage
%\footnotesize

\end{document}